\documentclass[12pt]{iopart}
\usepackage[latin9]{inputenc}
\usepackage{color}
\usepackage{float}
\usepackage{textcomp}
\usepackage{amsmath}
\usepackage{amssymb}
\usepackage{graphicx}
\usepackage{esint}

\makeatletter
\usepackage{iopams}
\usepackage{setstack}

\makeatother

\begin{document}

\title{Transient Charge and Energy Balance in Graphene Induced by Ultrafast
Photoexcitation}

\author{Junhua Zhang,$^{1}$ J$\ddot{o}$rg Schmalian,$^{2}$ Tianqi Li,$^{3}$
and Jigang Wang$^{3}$}

\address{$^{1}$Department of Physics, College of William and Mary, Williamsburg,
Virginia 23187, USA\\
 $^{2}$Institute for Theory of Condensed Matter and Center for Functional
Nanostructures, Karlsruhe Institute of Technology, Karlsruhe 76128,
Germany\\
 $^{3}$Ames Lab and Department of Physics and Astronomy, Iowa State
University, Ames, Iowa 50011, USA}
\begin{abstract}
Ultrafast optical pump-probe spectroscopy measurement on monolayer
graphene observes significant optical nonlinearities. We show that
strongly photoexcited graphene monolayers with $35$fs pulses \emph{quasi-instantaneously}
build up a \emph{broadband, inverted} Dirac fermion population. Optical
gain emerges and directly manifests itself via a negative conductivity
at the near-infrared region for the first 200fs, where stimulated
emission completely compensates absorption loss in the graphene layer.
To quantitatively investigate this transient, extremely dense photoexcited
Dirac-fermion state, we construct a two-chemical-potential model,
in addition to a time-dependent transient carrier temperature above
lattice temperature, to describe the population inverted electronic
state metastable on the time scale of tens of femtoseconds generated
by a strong exciting pulse. The calculated transient optical conductivity
reveals a complete bleaching of absorption, which sets the saturation
density during the pulse propagation. Particularly, the model calculation
reproduces the negative optical conductivity at lower frequencies
in the states close to saturation, corroborating the observed femtosecond
stimulated emission and optical gain in the wide near-infrared window.
\end{abstract}
\maketitle

\section{Introduction}

Despite the well-established linear optical properties in graphene,
which is marked by a universal absorption $A=\pi\alpha=2.3\%$ ranging
from near-infrared to visible light,\cite{Nair08,Mak08,CastroNeto09}
significantly less attention has been paid to the ultrafast nonlinear
optical properties. Important for future photonic and optoelectronic
applications,\cite{Bonaccorso10} carrier dynamics in graphene after
being driven far out of equilibrium needs to be understood. Still,
ultrafast spectroscopy studies are reported recently to show unusual
properties.\cite{Sun08,Choi09,Lui10,Liu10,Stohr10,Dawlaty08,George08,Breusing09,Newson09,JPCL12,Dani12,Ryzhii07,Tombet12,TianqiLi10,Winnerl13,Winzer12,BYSun1212}
Especially, the observation of nonlinear absorption when applying
an ultrashort intense laser pulse to monolayer graphene reveals an
extremely dense, quasithermal photoexcited-carrier state created by
strong pumping on 10 fs time scale and metastable for several tens
of femtoseconds, which implies a unique transient electronic state
in the important emerging material graphene.\cite{TianqiLi10}

\begin{figure*}
\hfill{}\includegraphics[scale=0.6]{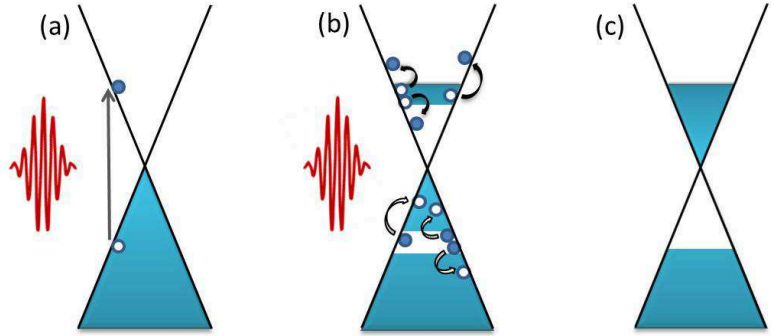}\hfill{} \caption{Schematic illustration of the formation of population inverted electronic
state in the intermediate time $\tau_{\text{th}}<t<\tau_{\text{c,r}}$.
(a) Photoexcited carriers generated by $\sim$10fs pump pulse; (b)
The leading scattering processes of photo-excited carriers taking
place in several femtoseconds: $e+e\rightarrow e+e,\ h+h\rightarrow h+h,\ e+h\rightarrow e+h$,
which quickly establish individual thermalization in electron and
hole carriers sharing a common electronic temperature $T_{e}$ due
to the electron-hole scattering events. (c) After the internal thermalization,
the photoexcited carriers form a population inverted hourglass-like
electronic state characterized by two chemical potentials and a common
electron temperature. \label{fig:Schematic-illustration}}
\end{figure*}

When strongly driven out of equilibrium by coherent light, the excited
carriers subsequently participate in several dynamical processes during
relaxing back to its equilibrium. Among them are the carrier decoherence,
thermalization, cooling, and electron-hole recombination. If the excitation
pulse is short enough, by observing the responses followed from right
after the pump, we can identify the typical time scales associated
with theses processes. Facilitated by recent ultrafast spectroscopy
measurements, certain progress has been made. It is recognized that
the ultrafast carrier dynamics in graphene is different from that
in metals and semiconductors. First of all, dimensional estimates
for Dirac fermions in graphene yield the carrier decoherence time
$\tau_{\text{dc}}^{-1}\sim\frac{e^{4}}{(\hbar v)^{2}}\hbar\omega$,\cite{Fritz08}
which becomes rather short, $\tau_{\text{dc}}\sim1\text{fs}$, for
pump photon energy on the order of 1eV. Rapid carrier-carrier scattering
gives rise to an electronic thermalization time $\tau_{\text{th}}$
on the order of 10 fs.\cite{Breusing09,TianqiLi10} Time-resolved
studies observe different cooling time scales with the shortest $\tau_{\text{c}}\sim100\text{fs}$
from electron-optical phonon coupling and longer electron-hole recombination
time $\tau_{\text{r}}$ on the ps time scale.\cite{Sun08,George08,Dawlaty08,Choi09,Breusing09,Newson09,Tse_DasSarma09,Bistritzer_MacDonald09,Winzer10}
Thus, when applying an ultrashort pump pulse of $\tau_{\text{p}}\sim10\text{fs}$,
the distinct time scales $\tau_{\text{th}}\ll\tau_{\text{c,r}}$ entails
an intermediate electronic state purely determined by carrier-carrier
scattering. Unlike in most semiconductors where $\tau_{\text{th}}\sim100\text{fs}>\tau_{\text{p}}$,
the extremely short decoherence time and the rapid carrier-carrier
scattering quickly deplete the phase space from the neighbourhood
of optical excitation to the whole band to fully thermalize the carriers
as illustrated in Fig. \ref{fig:Schematic-illustration} (b). The
analysis of Coulomb interaction between the Dirac-fermionic excitations
shows a slow population imbalance relaxation,\cite{Fritz08,Foster09}
in particular at high pulse intensity due to the suppression of Auger
processes for strong pulse excitation,\cite{Winzer10} although there
is no gap. Thus for $\tau_{\text{th}}<t<\tau_{\text{c,r}}$ the number
of photoexcited carriers in each branch of the Dirac cone decays rather
slowly. Therefore, the fast decoherence and thermalization together
with slow imbalance relaxation support a unique population-inverted
electronic state adiabatically formed after the strong pump excitation,
a quasi-stable ``hourglass'' state on the time scale of some tens
of femtoseconds, as shown in Fig. \ref{fig:Schematic-illustration}
(c).

\begin{figure}[H]
\hfill{}\includegraphics[scale=0.66]{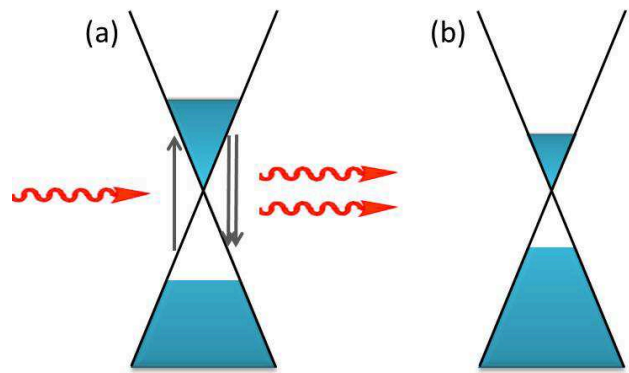}\hfill{}

\caption{Schematic illustration of (a) the femtosecond stimulated emission
and optical gain when applying a low-frequency pulse to a high-density
population-inverted state created by an intense high-frequency pump
and (b) the resulting metastable low-density state. \label{fig:stimulated emission}}
\end{figure}

In this article, we focus on this unique transient electronic state
and provide details on ultrafast optical pump-probe spectroscopy measurement
on monolayer graphene to reveal significant ultrafast optical nonlinearities,
including nonlinear absorption saturation and near-infrared stimulated
emission. These properties arise from a \emph{broadband, inverted}
Dirac fermion population induced by $35$fs pulse excitation. Optical
gain emerges and directly manifests itself via a negative conductivity
at the near-infrared region for the first 100s of fs, where stimulated
emission completely compensates absorption loss in the graphene layer.
To quantitatively investigate this transient, extremely dense photoexcited
Dirac-fermion state, we construct a simple model of a quasi-thermalized
distribution with one electron temperature but two distinct chemical
potentials associated with the electron- and hole-band, respectively.
We find this transient state associated with high electron temperature
$T_{e}$ up to 3000-4000K, which causes a broadband distribution extending
to high energy that naturally explains the observed blueshifted component
in photoluminescence spectrum.\cite{Lui10,Liu10,Stohr10} We further
explore the phase space capacity and identify the maximal photoexcitation
density restricted by phase space filling. and individual chemical
potential of each band are calculated. To understand the observed
nonlinear optical behavior and the measured large saturation density,
we calculate the optical conductivity for this nonequilibrium electronic
state. The results show that the available phase space cannot be completely
filled but will be saturated at a lower photoexcitation density due
to the balance between absorption and emission. This calculated saturation
density is in excellent agreement with the experimental value.\cite{TianqiLi10}
Most interestingly, our model reproduces the experimental results
that the nearly-saturated states created by a high-frequency pump
are unstable to a low-frequency pulse through stimulated emission
to bring the system to a lower-level metastable state (illustrated
in Fig. \ref{fig:stimulated emission}), resulting in an optical gain
phenomenon within the first 100s of fs after photoexcitation. The
excellent agreement between theory and experiment further corroborates
that our simple model captures the feature of the transient state
at early timescale ($<100\text{fs}$) in the high excitation regime.
Meanwhile, the comparison with the measured optical conductivity disfavors
the equal-chemical-potential model to describe the transient states
in graphene at the fs time scale, an outstanding issue debated in
the community.

\section{Experimental Details}

\subsection{Spectroscopy measurement}

\begin{figure}[H]
\hfill{}\includegraphics[scale=0.66]{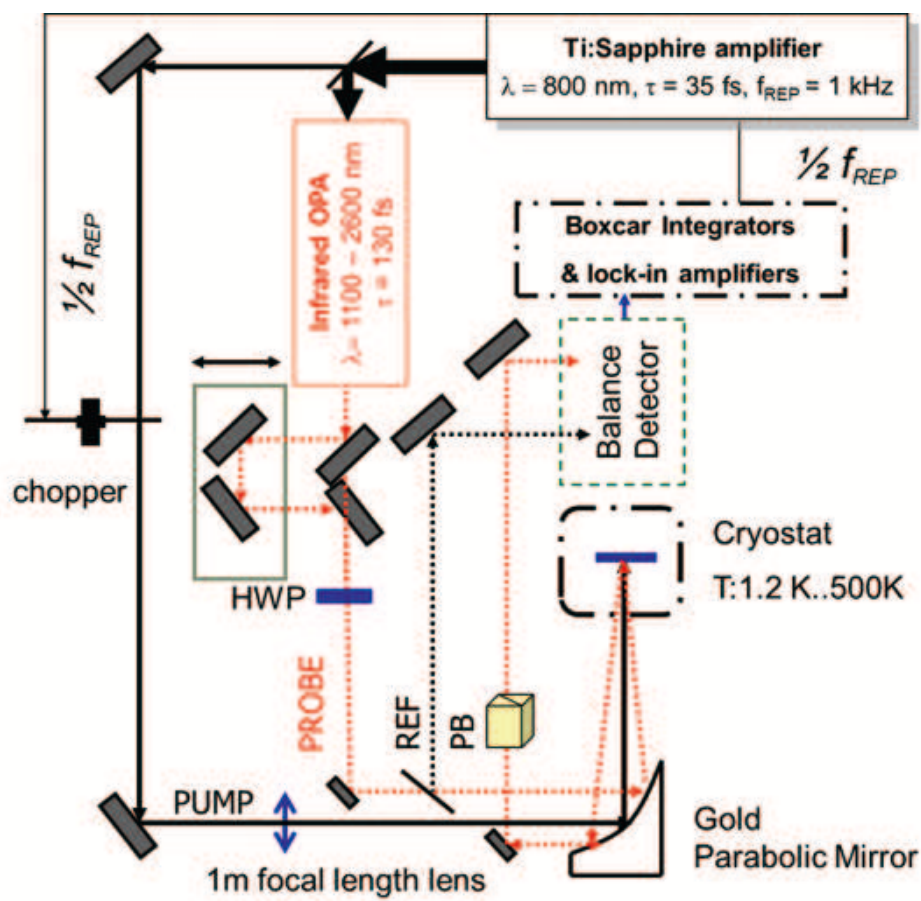}\hfill{}

\caption{Experimental Schematics. HWP: half waveplate; PB: polarizing beam
splitter, FL: focal length.}

\label{Fig1} 
\end{figure}

The experimental setup is shown in Fig. \ref{Fig1}. In our experiment,
the Ti:Sapphire amplifier with center wavelength 800nm, pulse width
35fs at 1kHz repetition rate is used. This further drives an optical
parametric amplifiers tunable with tunable optical pump pulses covering
572-2400 nm allowing for both degenerate and non-degenerate pump/probe
spectroscopy. The laser is further split into pump and probe paths.
The pump beam, chopped as half harmonic of the laser repetition rate,
directly excites the sample. The reflection of the probe beam, together
with reference, is fed into an auto-balance detector, and the individual
beams as well as the difference between them are picked up by three
boxcar integrators. During the measurement, the pump fluence from
few $\mu$J$/cm^{2}$ to $m$J$/cm^{2}$ level is finely controlled.
This way we can record pump-induced differential reflectivity changes
$\Delta R/R$ with $\sim$40fs time resolution and signal-to-noise
ratio down to 5$\cdot$10$^{-5}$. Similar experimental setups and
details are described elsewhere, e.g., see Refs. \cite{Wang04,Sanders05,Wang09,Wang10,Wang042,LiNature13}.

\subsection{Samples}

Graphene was prepared from the thermal evaporation of SiC \cite{Hupalo09}
with substrates used in the current experiments 6H-SiC(0001) purchased
from Cree, Inc.. The samples were graphitized in UHV ($\mathrm{P}\sim1\times10^{-10}$torr)
by direct current heating of the sample to $\sim$1300\textdegree{}C
measured with an infrared pyrometer with reading of the pyrometer
adjusted to take account of the graphite emissivity reported in the
literature \cite{Forbeaux02}. The sample was not pretreated in a
H2 atmosphere within a furnace which is a common practice because
it excludes the formation of multi-step heights and easier control
of thickness. The layer thickness (whether single layer G1 or bilayer
G2) was controlled by the heating rate: faster one-step heating rates
(within 2-3 seconds to reach 1300\textdegree{}C) result in large G1
domains while multiple heating steps with a slower rate (30seconds
to reach 1350\textdegree{}C) result in samples with large G2 areas.
Graphene thickness was identified using contrast thickness \cite{Riedl07}
and with step heights changes between different regions which were
found to be combinations of only two steps ,i.e., 0.25nm (of SiC),
and 0.33nm (of graphene ) as explained in Ref.\cite{Hupalo09}. Fig.
1c in Ref.\,\cite{TianqiLi10} shows a large $2\mu\mathrm{m}\times2\mu\mathrm{m}$
(left) G1 formed after heating with the fast rate. The atomic scale
image is shown to the right with the $1\times1$ unit cell seen with
lattice constant 0.246nm and intensity modulation due to the $6\sqrt{3}$
is also seen. The tunneling conditions are -0.5V, 1nA. The high intensity
of the modulation and the resolution of the 6 atoms of the graphene
ring indicate that this is predominantly G1(in excess of 90\% of the
area). The detailed growth conditions, characterizations and doping
($\sim$0.4 eV for monolayer) of the obtained epitaxial graphene on
SiC are extensively established by our papers \cite{Hupalo09,HupaloAM2011}
and many others in the literature, e.g., \cite{GierzNL2008,Zhou07,GierzPRB2010}.

\section{Threshold reflection coefficient and optical gain}

Considering graphene on a substrate with dielectric constant $\varepsilon_{s}$,
the amplitude of the reflected and transmitted waves for a normal
incident beam follow from Maxwell's equations along with the usual
boundary conditions: 
\begin{eqnarray}
\hat{r} & = & \frac{1-n_{s}-4\pi\sigma\left(\omega\right)/c}{1+n_{s}+4\pi\sigma\left(\omega\right)/c},\nonumber \\
\hat{t} & = & \frac{2}{1+n_{s}+4\pi\sigma\left(\omega\right)/c}.\label{amplitudes}
\end{eqnarray}
Here $\sigma\left(\omega\right)$ is the complex optical conductivity.
The common reflection and transmission coefficients are determined
by $R=\left\vert \hat{r}\right\vert ^{2}$ and $T=n_{s}\left\vert \hat{t}\right\vert ^{2}$.
In case $\sigma=0$ holds that $R+T=1$. The presence of a finite
conductivity in the graphene sheet leads to absorption 
\begin{equation}
A_{g}=\frac{1}{4}\left(1+n_{s}\right)^{2}\left(1-T-R\right).\label{absorption}
\end{equation}
where it is custom\cite{MakPRL2008} to introduce the coefficient
$\left(1+n_{s}\right)^{2}/4$ such that $A_{g}$ corresponds to the
absorption coefficient of a suspended graphene sheet.

Following Ref.\cite{MakPRL2008} we can introduce the reflection of
the substrate (for $\sigma=0$) 
\begin{equation}
R_{s}=\left(\frac{1-n_{s}}{1+n_{s}}\right)^{2}
\end{equation}
and of the substrate with graphene $R_{s+g}$ 
\begin{equation}
R_{s+g}=\left\vert \frac{1-n_{s}-4\pi\sigma\left(\omega\right)/c}{1+n_{s}+4\pi\sigma\left(\omega\right)/c}\right\vert ^{2}.
\end{equation}
For the complex optical conductivity of graphene in equilibrium and
at $T=0$ it holds 
\begin{equation}
\sigma_{eq}\left(\omega,T=0\right)=\frac{e^{2}}{4\hbar}\left(\theta\left(\omega-2\mu\right)-\frac{i}{2\pi}\ln\left(\frac{\omega+2\mu}{\omega-2\mu}\right)^{2}\right).
\end{equation}
Near the jump in the optical conductivity at $\omega=2\mu$, the imaginary
part of the conductivity has a logarithmic divergence which is smeared
out in case of finite temperatures. Since $\sigma\left(\omega\right)$
is of order $e^{2}/\hbar$, it holds that $\sigma\left(\omega\right)/c$
is of order of the finestructure constant of quantum electrodynamics
$\alpha_{\mathrm{QED}}=e^{2}/\left(\hbar c\right)\approx1/137\ll1$.
This allows for an expansion in $\sigma\left(\omega\right)/c$. It
follows 
\begin{equation}
\frac{R_{s+g}-R_{s}}{R_{s}}=\frac{4}{n_{s}^{2}-1}\frac{4\pi}{c}\sigma^{\prime}\left(\omega\right)+O\left(\alpha_{\mathrm{QED}}^{2}\right)\label{refl}
\end{equation}
Thus, the reflection coefficient to leading order in $\alpha_{\mathrm{QED}}$
is fully determined by the real part of the optical conductivity $\sigma^{\prime}\left(\omega\right)=\mathrm{Re}\sigma\left(\omega\right)$,
the imaginary part only enters at higher orders. For the transmission
and absorption coefficients follows in the same limit 
\begin{eqnarray}
T & = & \frac{4n_{s}}{\left(1+n_{s}\right)^{2}}-\frac{8n_{s}}{\left(n_{s}+1\right)^{3}}\frac{4\pi}{c}\sigma^{\prime}\left(\omega\right)+O\left(\alpha_{\mathrm{QED}}^{2}\right)\nonumber \\
A_{g} & = & \frac{4\pi}{c}\sigma^{\prime}\left(\omega\right)+O\left(\alpha_{\mathrm{QED}}^{2}\right).\label{tem}
\end{eqnarray}
This yields the result 
\begin{equation}
\frac{R_{g+s}-R_{s}}{R_{s}}=\frac{4}{n_{s}^{2}-1}A_{g}
\end{equation}
of Ref.\cite{MakPRL2008}.

Eq.\ref{refl} enables us to determine a threshold value for the reflectivity
that corresponds to a negative optical conductivity and thus to a
behavior with optical gain. From Eq.\ref{refl} it follows for the
reflection after delay time $\tau$: 
\begin{eqnarray}
\Delta R/R & \equiv & \frac{R_{s+g}\left(\tau\right)-R_{s+g}\left(0\right)}{R_{s}}\nonumber \\
 & = & \frac{4}{n_{s}^{2}-1}\frac{4\pi}{c}\left(\sigma^{\prime}\left(\tau\right)-\sigma^{\prime}\left(0\right)\right)
\end{eqnarray}
Using the experimentally established value $\sigma^{\prime}\left(0\right)=e^{2}/\left(4\hbar\right)$
for the optical conductivity prior to the pulse, it follows that $\sigma^{\prime}\left(\tau\right)<0$
if $\Delta R/R<\left.\Delta R/R\right\vert _{c}$ where 
\begin{equation}
\left.\Delta R/R\right\vert _{c}=-\frac{4\pi\alpha_{\mathrm{QED}}}{n_{s}^{2}-1}.
\end{equation}
With $n_{s}=2.7$ it follows $\left.\Delta R/R\right\vert _{c}=-1.4582\%.$
If for some reason the dielectric constant of the substrate is larger
that $2.7$, this would only reduce the critical value of $\Delta R/R$
and we would only underestimate the regime where $\sigma<0$. Given
that our data yield the magnitude of $\Delta R/R$ as big as $1.9\%$,
it follows that we have $\sigma<0$ as long as $n_{s}>2.41$. In the
literature, the uncertainty of $n_{s}=2.7$ is $\pm0.1$. The smallest
index of SiC is $2.55$ in the THz range. These results demonstrate
that our conclusion $\sigma<0$ is robust.

\begin{figure}[H]
\hfill{}\includegraphics[scale=0.6]{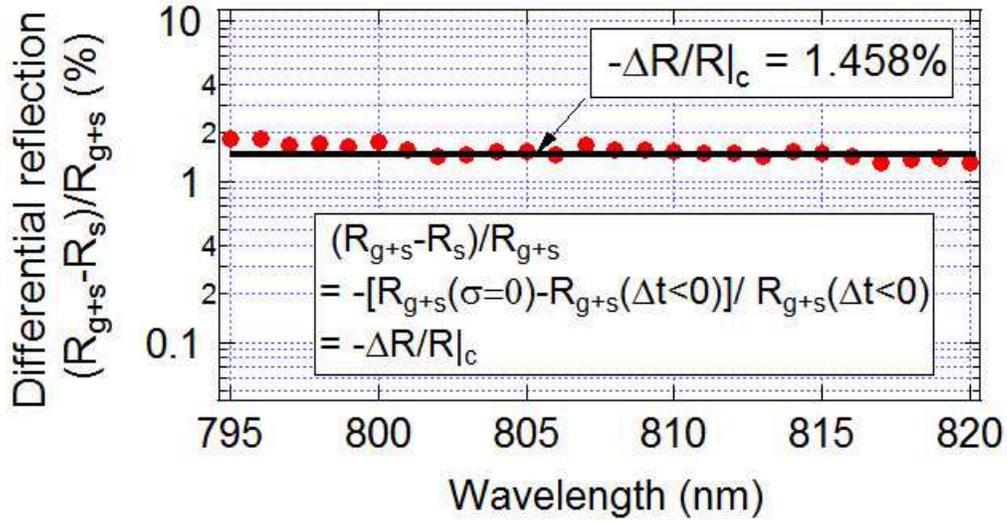}\hfill{}

\caption{The differential reflectivity determined by the measurements with
R$_{g+s}$ or without R$_{s}$ graphene monolayer on SiC substrate.
The reflection from the zero conductivity in pumped garphene/SiC exactly
corresponds to the case of bare SiC substrate. Consequently, the threshold
$\left.\Delta R/R\right\vert _{c}$ for zero conductivity can be directly
determined from the curve, which is consistent with value used in
the manuscript $\sim$-1.46$\%$. }

\label{fig: thres} 
\end{figure}

In Fig. \ref{fig: thres}, we experimentally determine the existence
and value of the threshold $\left.\Delta R/R\right\vert _{c}=-1.4582\%$
for zero conductivity in our sample. This further demonstrates unambiguously
that the reflectivity geometry in current sample provides a direct
measurement of the real part of conductivity $\sigma$ of the graphene
layer (or absorption), which directly accessing the gain/loss processes.
This also demonstrates again our sample is graphene monolayer, consistent
with conclusion from STM.

Using the same reasoning we can relate the reflectivity to the absorption
coefficient 
\begin{equation}
R_{s+g}\left(\tau\right)=R_{s}+\frac{n_{s}-1}{n_{s}+1}\frac{4}{\left(1+n_{s}\right)^{2}}A\left(\tau\right)
\end{equation}
and obtain 
\begin{equation}
\frac{A_{g}\left(\tau\right)-A_{g,0}}{A_{g,0}}=\frac{R_{s+g}\left(\tau\right)-R_{s+g}\left(0\right)}{R_{s+g}\left(0\right)}\cdot\frac{n_{s}^{2}-1+4A_{g,0}}{4A_{g,0}}\label{eq: delta A over A}
\end{equation}
that will be used in our analysis of the density of transient electrons
and holes, where $A_{g,0}=A_{g}\left(\tau=0\right)=\pi\alpha_{\mathrm{QED}}$.

\section{Stimulated infrared emission and optical gain}

\begin{figure}[H]
\hfill{}\includegraphics[scale=0.66]{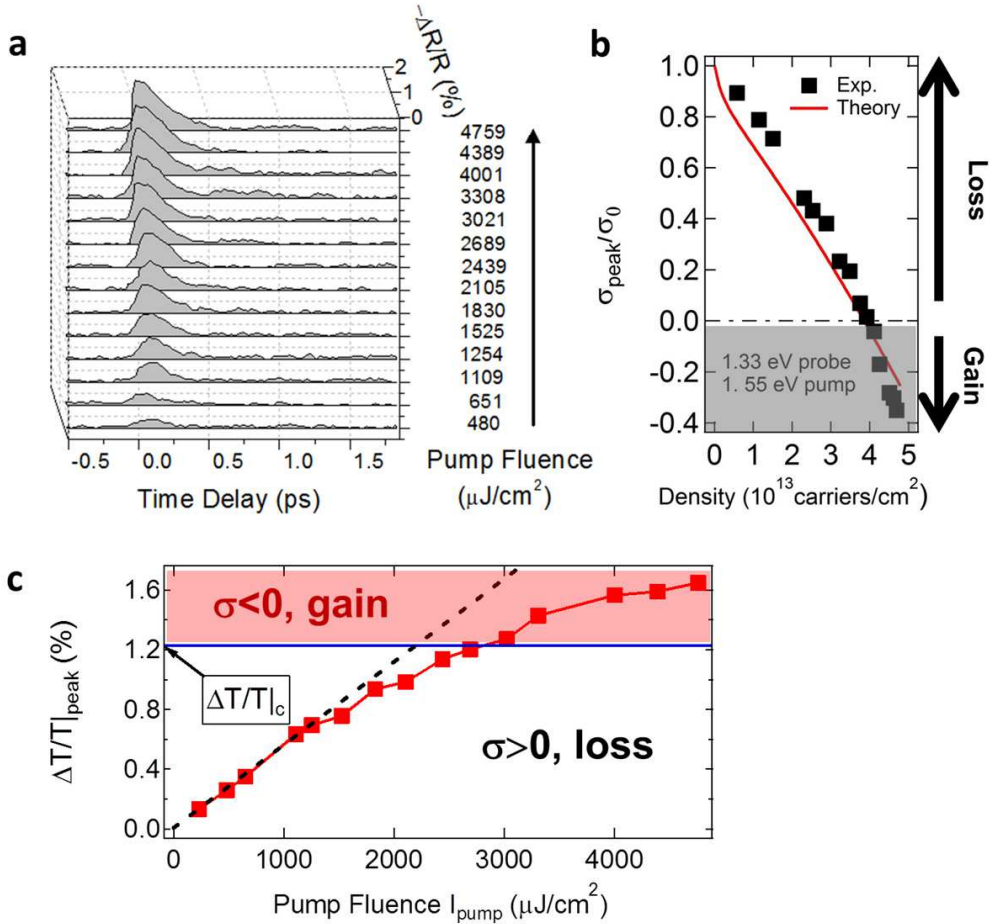}\hfill{}

\caption{(a) Ultrafast $\Delta R/R$ at 1.55 eV pump and 1.33 eV probe for
different pump fuences from 480 to 4759$\mu$J$/cm^{2}$(b) Experimental
values (rectangles) vs. theory (line) for peak transient conductivity,
showing negative conductivity above a threshold pump fluence. (c)
The extracted peak transient transmission as function of the pump
fluence clearly shows the positive transmission change, nonlinear
saturation, and that the critical vaule $\frac{\Delta T}{T}|_{c}$
(blue line) for zero conductivity indeed occurs. }

\label{fig: 930nm} 
\end{figure}

Here we provide a set of pump fluence dependence data at probe photon
probe energy at 1.33 eV, as shown in Figs. \ref{fig: 930nm}a and
\ref{fig: 930nm}b. Our conclusions w.r.t. stimulated emission and
optical gain are based on the observed negative conductivity in strongly
photoexcited graphene, which is fully consistent with the complementary
data presented in Ref \cite{TianqiLi10}. In addition, following the
similar analysis Eqs. (6) and (7), the differential transmission of
our sample can be extracted by the information of the differential
reflectivity or the subsequently derived conductivity of the graphene
sample. More importantly, there also exists a threshold value for
the photoinduced differential transmission $\frac{\Delta T}{T}|_{c}$
that corresponds to a zero optical conductivity, above which the optical
gain has to emerge because of the negative conductivity. 
\begin{equation}
\frac{\Delta T}{T}|_{c}=\frac{2\pi\alpha_{QED}}{n_{s}+1}
\end{equation}

From Eq.\ref{refl} and Eq.\ref{tem} it follows for the differential
transmission after delay time $\tau$: 
\begin{equation}
\frac{\Delta T}{T}(\tau)=\frac{\Delta R}{R}(\tau)\cdot\frac{1-n_{s}}{2}.
\end{equation}
The extracted peak transient transmission as function of the pump
Fluence, as shown in Fig. \ref{fig: 930nm}c, clearly shows the positive
transmission change and, mostly critically, that the critical value
(blue line) for zero conductivity indeed occurs. While those data
for the transmission were obtained indirectly, from our reflectivity
measurements, a direct measurement of the transmission would be an
important confirmation of our results ideally using large area free
standing graphene monolayer samples.

\section{Analysis of the density of transient electrons/holes}

\begin{figure}[H]
\hfill{}\includegraphics[scale=0.5]{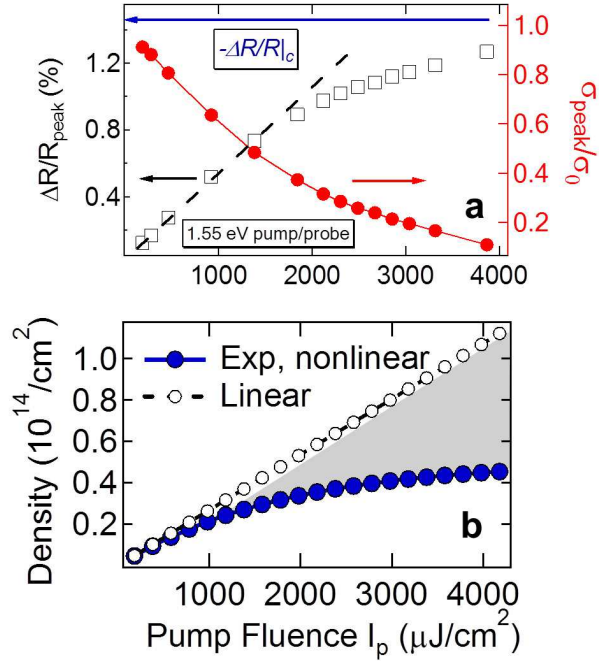}\hfill{} \caption{(a) The peak ${\frac{\triangle R}{R}}|_{peak}$ as function of the
pump fluence (black squares) measured by degenerate differential reflectivity
at 1.55 eV for the graphene monolayer and the corresponding conductivity
change (red solid dots). Blue arrow marks the threshold for zero conductivity
$\left.\Delta R/R\right\vert _{c}=-1.4582\%$ (see text). Dashed line:
linear dependence (guide to the eyes). (b) The extracted transient
fermion density at 40 fs (blue dots), as explained in the text, which
is significantly lower than $A_{0}I_{p}/\hbar\omega$ obtained from
the universal conductivity (open circles), as illustrated in shadow
area. }

\label{fig} 
\end{figure}

The amplitude of the time dependent absorption $A$ as function of
pump fluence can be derived from the measured differential reflectivity
by applying the Fresnel equations in thin film limit \cite{MakPRL2008,book}
\begin{equation}
\frac{\Delta A(I_{p})}{A_{0}}=\frac{\Delta R_{g+s}(I_{p})}{R_{g+s}}\cdot\frac{n_{s}^{2}-1+4A_{g,0}}{4A_{g,0}},\label{eq: Fresnel}
\end{equation}
where $R_{g+s}$ and $\Delta R_{g+s}$ are the static reflectivity
and pump-induced reflectivity changes for the graphene monolayer (g)
on the substrate (s) with index $n_{s}=2.7$. $A_{g,0}$ is the absorption
of graphene monolayer without pump, which takes a universal value
of $A_{g,0}=\pi\frac{e^{2}}{\hbar c}\simeq0.023$, as determined by
the universal a.c. conductivity $\sigma_{0}=\frac{e^{2}}{4\hbar}$.
Here the $\Delta A(I_{p})/A_{0}$ is the relative differential absorption
of graphene on the substrate: $A_{0}=\frac{4}{(n_{s}+1)^{2}}A_{g,0}$,
which yields $\frac{\Delta A}{A_{0}}=\frac{\Delta A_{g}}{A_{g,0}}$.
Therefore Eq.(\ref{eq: Fresnel}) follows from Eq.(\ref{eq: delta A over A}).
The peak amplitude $A(I_{p})=A_{0}+\Delta A(I_{p})$ gradually diminishes
as increasing the pump fluence. From the measured transient saturation
curve above, one can extract the density of photoexcited electrons(holes)
in graphene after the propagation of a single laser pulse of 35 fs
($\tau_{p}$) with pump fluence $I_{p}$ 
\begin{equation}
n_{\text{ex}}(I_{p})=\int_{-\infty}^{\infty}\frac{\mathrm{d}t}{\tau_{p}}n_{\text{ex}}(t,I_{p})=\frac{1}{\hbar\omega}\int_{-\infty}^{\infty}\frac{\mathrm{d}t}{\tau_{p}}I(t,I_{p})A\left(t\right),\label{eq: eq2}
\end{equation}
where $I(t,I_{p})$ is the Gaussian pulse envelop $I(t,I_{p})=I_{p}\sqrt{\frac{4\ln2}{\pi}}\exp\left[\frac{-4\ln2}{\tau_{p}^{2}}t^{2}\right]$,
normalized such that the total pulse fluence is $I_{p}=\int_{-\infty}^{\infty}\frac{\mathrm{d}t}{\tau_{p}}I(t,I_{p})$.
Since $A(t)=A_{0}+\Delta A(t)=A_{0}(1+\frac{\Delta A(t)}{A_{0}})$,
we have 
\begin{equation}
n_{\text{ex}}(t,I_{p})=\frac{I(t,I_{p})A_{0}}{\hbar\omega}\left(1+\frac{\Delta A(t)}{A_{0}}\right).\label{density}
\end{equation}
Applied to graphene where $\tau_{\mathrm{th}}\ll\tau_{p}$, $A\left(t\right)$
is determined by the adiabatic dependence of the absorption on the
pump fluence with $I_{\text{partial}}\left(t,I_{p}\right)=\int_{-\infty}^{t}\frac{\mathrm{d}t^{\prime}}{\tau_{p}}I(t^{\prime},I_{p})$.
Consequently, Eq.\ref{density} becomes 
\begin{equation}
n_{\text{ex}}(t,I_{p})=\frac{I(t,I_{p})A_{0}}{\hbar\omega}\left(1+\frac{\Delta A\left(I_{\text{partial}}(t,I_{p})\right)}{A_{0}}\right).\label{eq: eq4}
\end{equation}
We determine $A\left(I_{\text{partial}}\right)$ experimentally from
the reflectivity data of Fig. \ref{fig}a, combined with Eq.\ref{eq: Fresnel},
as discussed above. Finally, from Eqs. (\ref{eq: Fresnel})-(\ref{eq: eq4})
we have 
\begin{equation}
n_{\text{ex}}(I_{p})=\int_{-\infty}^{\infty}\frac{I(t,I_{p})A_{0}}{\hbar\omega}\times\left[1+\frac{\Delta R_{g+s}\left(I_{\text{partial}}(t,I_{p})\right)}{R_{g+s}}\cdot\frac{n_{s}^{2}-1+4A_{g,0}}{4A_{g,0}}\right]\frac{\mathrm{d}t}{\tau_{p}}.
\end{equation}
The result is shown in Fig. \ref{fig}b, which clearly shows that
using the actual absorption $A\left(t\right)$, instead of $A_{0}$,
is crutial to understand the high density regime of fs dynamics in
graphene discussed here.

\section{Theory}

\subsection{Model}

We construct the theoretical model based on the following considerations:
the above analysis of the time scales associated with different dynamical
processes shows that the internal thermalization time $\tau_{\text{th}}$
of the electron system is much shorter than the cooling $\tau_{\text{c}}$
and the electron-hole recombination time $\tau_{\text{r}}$; Simultaneous
conservation of momentum and energy for Dirac-fermion scattering indicates
the leading scattering processes are $e+e\rightarrow e+e,\ h+h\rightarrow h+h,\ e+h\rightarrow e+h$
in strong excitation regime with suppression of Auger processes.\cite{Winzer10}
For the intermediate time regime $\tau_{\text{th}}<t<\tau_{\text{c,r}}$,
i.e., after absorbing photons but before losing energy to the lattice,
photoexcited electrons and holes in graphene quickly establish separate
thermal equilibrium through carrier-carrier scattering. This gives
rise to sharply separated chemical potentials for the two bands, similar
to the case of graphite thin film.\cite{Breusing09} Due to the scattering
between electron- and hole-carriers, $e+h\rightarrow e+h$, the whole
electronic system obtains a common temperature, the electron temperature
that differs from the lattice temperature. Therefore we characterize
this intermediate electronic state by two Fermi-Dirac distributions

\begin{equation}
f_{\pm}(k)=\frac{1}{\exp\left[\frac{\varepsilon_{\pm}(k)-\mu_{\pm}}{k_{\text{B}}T_{e}}\right]+1}\label{eq: distribution function}
\end{equation}
for the upper ($+$) and lower ($-$) bands of the Dirac spectrum
$\varepsilon_{\pm}(k)=\pm\hbar vk$ with separate chemical potential
$\mu_{\pm}$ but the same electron temperature $T_{e}$. Note that
for the convenience of theoretical calculation, we use the upper-
and lower-band electron picture instead of the electron-hole picture,
which are related via $f_{+}=f_{e},\ f_{-}=1-f_{h}$ and $\mu_{+}=\mu_{e},\ \mu_{-}=-\mu_{h}$.
To solve for electron temperature and chemical potentials, we take
into account the following conservation laws: 1) the total number
of electrons before and after pump excitation is the same, 2) a pseudo-conservation
law due to the slow population imbalance relaxation valid in strong
excitation regime: the photoexcited carrier number in the intermediate
state stays the same as that of right after the pump excitation, 3)
and the above described adiabatic process requires that the absorbed
photon energy is kept in the electron system until the formation of
the quasi-thermal distribution (\ref{eq: distribution function}).
These three conditions are expressed as

\begin{equation}
n_{\text{tot}}=n_{+}+n_{-}=n_{+}^{0}+n_{-}^{0},\label{eq: model Eq 1}
\end{equation}
\begin{equation}
n_{\text{ex}}=n_{+}-n_{+}^{0}=n_{-}^{0}-n_{-},\label{eq: model Eq 2}
\end{equation}

\begin{equation}
n_{\text{ex}}\hbar\omega=u-u^{0},\label{eq: model Eq 3}
\end{equation}
where $n_{\text{tot}}=N_{\text{tot}}/L^{2}$ represents the total
density of electrons in the system, $n_{\text{ex}}=N_{\text{ex}}/L^{2}$
refers to the density of photoexcited carriers, $n_{\pm}$ ($n_{\pm}^{0}$)
indicate the electron densities in the intermediate (initially equilibrium)
state, and $u$ ($u^{0}$) represents the intermediate (initial) energy
density of the whole electron system while $\hbar\omega$ is the pump
photon energy. Applying the distribution (\ref{eq: distribution function})
to Eqs. (\ref{eq: model Eq 1})-(\ref{eq: model Eq 3}) and taking
into account the valley and spin degeneracy in graphene, we obtain
the following expressions in terms of fugacities $z^{0}=e^{\frac{\mu^{0}}{k_{\text{B}}T_{e}^{0}}},\ z_{\pm}=e^{\frac{\mu_{\pm}}{k_{\text{B}}T_{e}}}$
with initial temperature $T_{e}^{0}=300\text{K}$, 
\begin{eqnarray}
\delta & = & \frac{g}{2\pi}\frac{(k_{\mathrm{B}}T_{e})^{2}}{(\hbar v)^{2}}\left[-\mathrm{Li}_{2}(-z_{+})+\mathrm{Li}_{2}\left(-\frac{1}{z_{-}}\right)\right]\nonumber \\
 & = & \frac{g}{2\pi}\frac{(k_{\mathrm{B}}T_{e}^{0})^{2}}{(\hbar v)^{2}}\left[-\mathrm{Li}_{2}(-z^{0})+\mathrm{Li}_{2}\left(-\frac{1}{z^{0}}\right)\right],\label{eq: concret Eq 1}
\end{eqnarray}
with $\delta$ referring to the initial doping density with respect
to the neutrality point, and 
\begin{eqnarray}
n_{\text{ex}} & = & \frac{g}{2\pi}\frac{1}{(\hbar v)^{2}}\left\{ (k_{\mathrm{B}}T_{e})^{2}\left[-\mathrm{Li}_{2}(-z_{+})\right]-(k_{\mathrm{B}}T_{e}^{0})^{2}\left[-\mathrm{Li}_{2}(-z^{0})\right]\right\} \nonumber \\
 & = & \frac{g}{2\pi}\frac{1}{(\hbar v)^{2}}\left\{ (k_{\mathrm{B}}T_{e})^{2}\left[-\mathrm{Li}_{2}\left(-\frac{1}{z_{-}}\right)\right]-(k_{\mathrm{B}}T_{e}^{0})^{2}\left[-\mathrm{Li}_{2}\left(-\frac{1}{z^{0}}\right)\right]\right\} ,\label{eq: concret Eq 2}
\end{eqnarray}
as well as 
\begin{eqnarray}
n_{\text{ex}}\hbar\omega & = & \frac{g}{\pi}\frac{(k_{\mathrm{B}}T_{e})^{3}}{(\hbar v)^{2}}\left[-\mathrm{Li}_{3}\left(-z_{+}\right)-\mathrm{Li}_{3}\left(-\frac{1}{z_{-}}\right)\right]\nonumber \\
 &  & -\frac{g}{\pi}\frac{(k_{\mathrm{B}}T_{e}^{0})^{3}}{(\hbar v)^{2}}\left[-\mathrm{Li}_{3}\left(-z^{0}\right)-\mathrm{Li}_{3}\left(-\frac{1}{z^{0}}\right)\right].\label{eq: concret Eq 3}
\end{eqnarray}
where $g=4$ is the flavor index taking into account the valley and
spin degeneracies, $v$ represents the Fermi velocity $v\approx1.1\times10^{6}$m/s
($\frac{1}{300}c$) in graphene, and the polylogarithm is defined
by a power series $\mathrm{Li}_{s}(z)=\sum_{n=1}^{\infty}\frac{z^{n}}{n^{s}}$.
Solving the three equations gives the transient electron temperature
$T_{e}$ and the individual chemical potentials $\mu_{\pm}=k_{\text{B}}T_{e}\ln z_{\pm}$
at a given photoexcitation density $n_{\text{ex}}$ with initial temperature
$T_{e}^{0}$ and initial chemical potential $\mu^{0}$ associated
with the equilibrium state before being excited.

To perform numerical calculation, we introduce the dimensionless variables
\begin{equation}
f_{\text{ex}}\equiv\frac{n_{\text{ex}}}{\bar{n}},\ \ x\equiv\frac{\delta}{\bar{n}},\ \ t_{e}\equiv\frac{k_{\text{B}}T_{e}}{D},\ \ \alpha_{\pm}\equiv\frac{\mu_{\pm}}{D},\ \ \Omega\equiv\frac{\hbar\omega}{D},\label{eq: dimensionless variables}
\end{equation}
with a choice for the upper momentum cutoff $\Lambda$ to define the
energy scale $D=\hbar v\Lambda$ and the density scale $\bar{n}=\frac{\Lambda^{2}}{\pi}$.
Here we choose $\Lambda$ such that $\pi\Lambda^{2}=\frac{1}{2}(2\pi)^{2}/A_{0}$
where $A_{0}=3^{3/2}a_{0}^{2}/2$ is the area of the hexagonal unit
cell. Note that these dimensionless units are solely introduced for
computational convenience. None of our final expressions depends on
the actual values of $\Lambda$, $D$ or $\bar{n}$, as these quantities
cancel in the final results (see for example Eq. (\ref{eq: maximum photoexcitation density})).
In terms of the dimensionless variables, the equations are expressed
as 
\begin{eqnarray}
f_{\mathrm{tot}}-1 & = & x\nonumber \\
 & = & \frac{g}{2}t_{e}^{2}\left[-\mathrm{Li}_{2}(-z_{+})+\mathrm{Li}_{2}\left(-\frac{1}{z_{-}}\right)\right]\nonumber \\
 & = & \frac{g}{2}\left(t_{e}^{0}\right)^{2}\left[-\mathrm{Li}_{2}(-z^{0})+\mathrm{Li}_{2}\left(-\frac{1}{z^{0}}\right)\right],\label{eq: dimensionless Eq 1}
\end{eqnarray}
and 
\begin{eqnarray}
f_{\text{ex}} & = & \frac{g}{2}\left\{ t_{e}^{2}\left[-\mathrm{Li}_{2}(-z_{+})\right]-\left(t_{e}^{0}\right)^{2}\left[-\mathrm{Li}_{2}(-z^{0})\right]\right\} \nonumber \\
 & = & \frac{g}{2}\left\{ t_{e}^{2}\left[-\mathrm{Li}_{2}\left(-\frac{1}{z_{-}}\right)\right]-\left(t_{e}^{0}\right)^{2}\left[-\mathrm{Li}_{2}\left(-\frac{1}{z^{0}}\right)\right]\right\} ,\label{eq: dimensionless Eq 2}
\end{eqnarray}
as well as 
\begin{eqnarray}
f_{\text{ex}} & = & \frac{g}{\Omega}\Biggl\{ t_{e}^{3}\left[-\mathrm{Li}_{3}\left(-z_{+}\right)-\mathrm{Li}_{3}\left(-\frac{1}{z_{-}}\right)\right]\nonumber \\
 &  & -\left(t_{e}^{0}\right)^{3}\left[-\mathrm{Li}_{3}\left(-z^{0}\right)-\mathrm{Li}_{3}\left(-\frac{1}{z^{0}}\right)\right]\Biggl\}.\label{eq: dimensionless Eq 3}
\end{eqnarray}
In the following analysis, either equation set (\ref{eq: concret Eq 1})-(\ref{eq: concret Eq 3})
or the set (\ref{eq: dimensionless Eq 1})-(\ref{eq: dimensionless Eq 3})
will be employed for convenience.

\subsection{Characteristics of the Intermediate Electronic State}

We carry out our analysis for two cases: undoped graphene, i.e. the
system at the charge neutrality point, and graphene on SiC substrate
with a finite electron doping. It is easy to show that the hole-doped
system is symmetric to the electron-doped system. By solving for the
electron temperature $T_{e}$ and the chemical potentials $\mu_{\pm}$
at different photoexcitation densities, we demonstrate the characteristics
of this intermediate electronic state.

\subsubsection{Neutral System}

The simplest case is the system at the neutrality point, i.e. $\delta=0$,
possessing particle-hole symmetry. Equation (\ref{eq: dimensionless Eq 1})
yields $z^{0}=1$ ($\mu^{0}=0$) and $z_{+}=\frac{1}{z_{-}}\equiv z$
($\mu_{+}=-\mu_{-}\equiv\mu$), i.e., the lower-band chemical potential
is always the opposite of the upper-band one in the neutral system.
From Eq. (\ref{eq: dimensionless Eq 2}) and (\ref{eq: dimensionless Eq 3})
we obtain the expression for the dimensionless temperature $t_{e}$
\begin{equation}
t_{e}=\left(\frac{\frac{f_{\mathrm{ex}}}{2}+\left(t_{e}^{0}\right)^{2}\frac{\pi^{2}}{12}}{-\mathrm{Li}_{2}(-z)}\right)^{1/2},\label{eq: neutral t}
\end{equation}
and the relation 
\begin{equation}
h(z)=\frac{\left(\frac{f_{\mathrm{ex}}}{2}+\left(t_{e}^{0}\right)^{2}\frac{\pi^{2}}{12}\right)^{3/2}}{\frac{f_{\mathrm{ex}}\Omega}{8}+\left(t_{e}^{0}\right)^{3}\frac{3}{4}\zeta(3)}\label{eq: neutral solving equation}
\end{equation}
with 
\begin{equation}
h(z)\equiv\frac{\left[-\mathrm{Li}_{2}(-z)\right]^{3/2}}{-\mathrm{Li}_{3}\left(-z\right)}.\label{eq: h(z)}
\end{equation}
Since $h(z)$ is monotonously increasing with an upper bound $3/\sqrt{2}$
in large $z$ limit, it implies a maximum value of $f_{\text{ex}}$:
$f_{\text{ex}}^{\text{max}}=\left(\frac{3\Omega}{4}\right)^{2}.$
That is to say, there exists a phase space limit on the photoexcited
carrier number:

\begin{equation}
n_{\text{ex}}^{\text{max}}=\frac{9}{16\pi v^{2}}\omega^{2}.\label{eq: maximum photoexcitation density}
\end{equation}
For instance, at $\hbar\omega=1.55\text{eV}$ the phase space capacity
becomes $n_{\text{ex}}^{\text{max}}=9.7779\times10^{13}\mathrm{cm}^{-2}$.
In this limit, the electron temperature approaches zero as $z\rightarrow\infty$,
$t_{e}\longrightarrow\frac{\left(f_{\text{ex}}^{\text{max}}\right)^{1/2}}{\ln z}=\frac{3\Omega}{4\ln z}\longrightarrow0.$

\begin{figure}[H]
\hfill{}\includegraphics[scale=0.8]{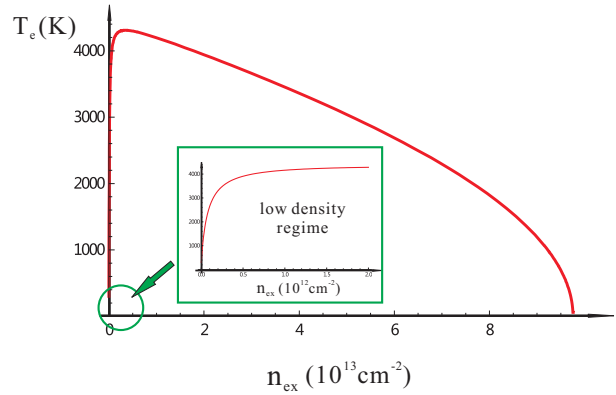}\hfill{}

\caption{Plot of the electron temperature $T_{e}$ varying with photoexcitation
density $n_{\text{ex}}$ in the neutral system. We can see that the
system is rapidly heated up at lower densities, but is slowly cooled
at higher densities. \label{fig: neutral system  T(n_ex)}}
\end{figure}

For a 800nm pump with photon energy $\hbar\omega=1.55\text{eV}$,
solving Eq. (\ref{eq: neutral t}) and (\ref{eq: neutral solving equation}),
we obtain $T_{e}$ as a function of $n_{\text{ex}}$ plotted in Fig.
\ref{fig: neutral system  T(n_ex)}. 

It shows that the electron temperature rises rapidly at low photoexcitation
densities, but, instead of keeping heated up, it starts slowly dropping
at higher densities and eventually approaches zero at a maximal density.

\begin{figure}[H]
\hfill{}\includegraphics[scale=0.8]{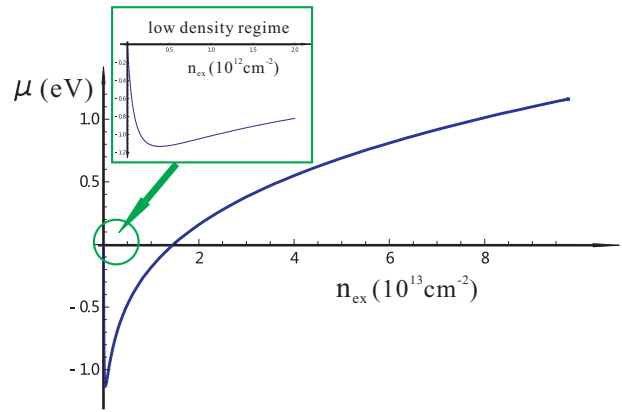}\hfill{}

\caption{Plot of the upper-band chemical potential $\mu_{+}(n_{\text{ex}})$,
in units of eV, in the neutral system. The lower-band chemical potential
$\mu_{-}=-\mu_{+}$. Clearly, the upper-band chemical potential turns
into negative at low densities before rising to positive at high densities.
\label{fig: neutral system MU(n_ex)}}
\end{figure}

The value of $\mu$ with respect to $n_{\text{ex}}$ is plotted in
Fig. \ref{fig: neutral system MU(n_ex)}. 

We clearly see that it turns into negative at low photoexcitation
densities during the rapidly heating up time, but back to positive
when temperature slowly decreasing.

The down-turn behavior in electron temperature and the negative-to-positive
transition in chemical potential signifies a crossover behavior that
at small pump fluence the excited carriers form a hot and dilute classical
gas, but with more carriers excited they gradually build up a quantum
degenerate fermion system with temperature cooling down in order to
accommodate more electrons in the finite phase space. If the phase
space could really be exhausted, the electron and the hole carriers
would be pumped into zero temperature Fermi-Dirac distributions in
which the carriers are closely packed with a sharp Fermi edge.

\subsubsection{Exhaustion of Phase Space}

\begin{figure}[H]
\hfill{}\includegraphics[scale=0.8]{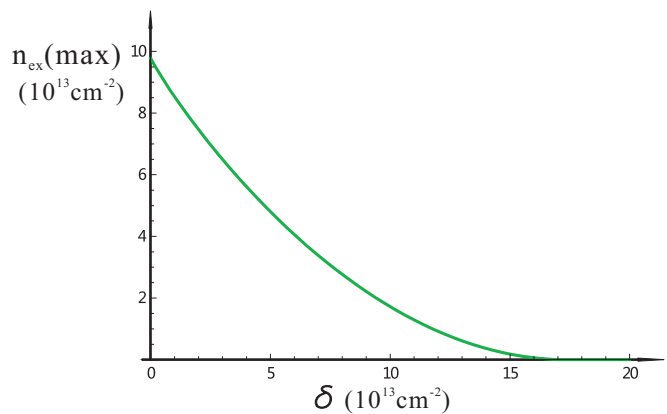}\hfill{}

\caption{Plot of the maximum photoexcited carrier density $n_{\text{ex}}^{\text{max}}$,
when the phase space is completely filled, as a function of initial
doping density $\delta$. The maximum photoexcitation density decreases
with increasing doping level, as expected. \label{fig: n_max vs doping}}
\end{figure}

Inspired by the analysis of the neutral system, we see that phase
space capacity is exhausted at zero electron temperature. To obtain
an analytical estimate of the maximal available phase space at different
electron-doping levels, we assume the initial temperature to be zero
for convenience. Equations (\ref{eq: model Eq 1})-(\ref{eq: model Eq 3})
are then simplified as 
\begin{equation}
\delta=\frac{1}{\pi}\frac{1}{\left(\hbar v\right)^{2}}\left[\left(\mu_{+}^{\text{max}}\right)^{2}-\left(\mu_{-}^{\text{max}}\right)^{2}\right]=\frac{1}{\pi}\frac{1}{\left(\hbar v\right)^{2}}\left(\mu^{0}\right)^{2},
\end{equation}

\begin{equation}
n_{\text{ex}}^{\text{max}}=\frac{1}{\pi}\frac{1}{\left(\hbar v\right)^{2}}\left[\left(\mu_{+}^{\text{max}}\right)^{2}-\left(\mu^{0}\right)^{2}\right]=\frac{1}{\pi}\frac{1}{\left(\hbar v\right)^{2}}\left(\mu_{-}^{\text{max}}\right)^{2},
\end{equation}

\begin{equation}
n_{\text{ex}}^{\text{max}}\hbar\omega=\frac{2}{3\pi}\frac{1}{\left(\hbar v\right)^{2}}\left[\left(\mu_{+}^{\text{max}}\right)^{3}+\left(-\mu_{-}^{\text{max}}\right)^{3}-\left(\mu^{0}\right)^{3}\right].
\end{equation}
In terms of the dimensionless variables defined in (\ref{eq: dimensionless variables}),
we find a relation between the maximal photoexcitation density and
the doping level from the above equations 
\begin{equation}
f_{\text{ex}}^{\text{max}}=\frac{2}{3\Omega}\left[\left(f_{\text{ex}}^{\text{max}}+x\right)^{3/2}+\left(f_{\text{ex}}^{\text{max}}\right)^{3/2}-x^{3/2}\right].\label{eq: fmax vs x}
\end{equation}
Solving this equation yields $n_{\text{ex}}^{\text{max}}$ at different
doping densities as shown in Fig. \ref{fig: n_max vs doping}. 

As seen from formula (\ref{eq: maximum photoexcitation density})
for the neutral system, $n_{\text{ex}}^{\text{max}}\sim\frac{\omega^{2}}{v^{2}}$,
the available phase space rises with increasing pump frequency. On
the other hand, phase space capacity decreases with increasing initial
electron doping density, as expected.

However, for photoexcitation density $n_{\text{ex}}$ as an input
parameter in our calculation, a critical question to ask is: does
this maximal density $n_{\text{ex}}^{\text{max}}$ equal the saturation
density $n_{\text{ex}}^{\text{sat}}$ in real pumping process, or
in another word, can phase space be completely filled? We will answer
this question later.

\subsubsection{Doped System}

Next we discuss the system away from the Dirac point with a finite
electron doping, i.e., $\delta\ge0$ or $x=\frac{\delta}{\bar{n}}\ge0$,
which is often the case, e.g., in epitaxial graphene on SiC substrate.
In this case, from Eqs. (\ref{eq: dimensionless Eq 1})-(\ref{eq: dimensionless Eq 3})
we obtain the expression of $t_{e}$ as a function of $z_{+}$ 
\begin{equation}
t_{e}=\left(\frac{\frac{f_{\text{ex}}}{2}+\left(t_{e}^{0}\right)^{2}\left[-\mathrm{Li}_{2}(-z^{0})\right]}{-\mathrm{Li}_{2}(-z_{+})}\right)^{1/2},\label{eq: Expt t}
\end{equation}
then the coupled equations are reduced to 
\begin{equation}
-\mathrm{Li}_{2}\left(-\frac{1}{z_{-}}\right)=C_{1}\left[-\mathrm{Li}_{2}(-z_{+})\right],\label{eq: General eq 1}
\end{equation}
\begin{equation}
\left[-\mathrm{Li}_{3}\left(-z_{+}\right)-\mathrm{Li}_{3}\left(-\frac{1}{z_{-}}\right)\right]^{2/3}=C_{2}\left[-\mathrm{Li}_{2}(-z_{+})\right],\label{eq: General eq 2}
\end{equation}
with 
\begin{equation}
C_{1}(f_{\text{ex}})=1-\frac{x}{f_{\text{ex}}+2\left(t_{e}^{0}\right)^{2}\left[-\mathrm{Li}_{2}(-z^{0})\right]},\label{eq: coefficient C1}
\end{equation}
\textcolor{black}{{} 
\begin{equation}
C_{2}(f_{\text{ex}})=\frac{\left(\frac{f_{\text{ex}}\Omega}{4}+\left(t_{e}^{0}\right)^{3}\left[-\mathrm{Li}_{3}\left(-z^{0}\right)-\mathrm{Li}_{3}\left(-\frac{1}{z^{0}}\right)\right]\right)^{2/3}}{\frac{f_{\text{ex}}}{2}+\left(t_{e}^{0}\right)^{2}\left[-\mathrm{Li}_{2}(-z^{0})\right]}.\label{eq: coefficient C2}
\end{equation}
}Solving the two equations (\ref{eq: General eq 1}) and (\ref{eq: General eq 2})
we obtain $z_{+}$ and $z_{-}$, which in turn gives $t$ via Eq.
(\ref{eq: Expt t}). Finally, the physical quantities are derived
through $k_{\text{B}}T_{e}=t_{e}D,\ \mu_{+}=t_{e}D\ln z_{+},\ \mu_{-}=t_{e}D\ln z_{-}.$

\begin{figure}[H]
\hfill{}\includegraphics[scale=0.8]{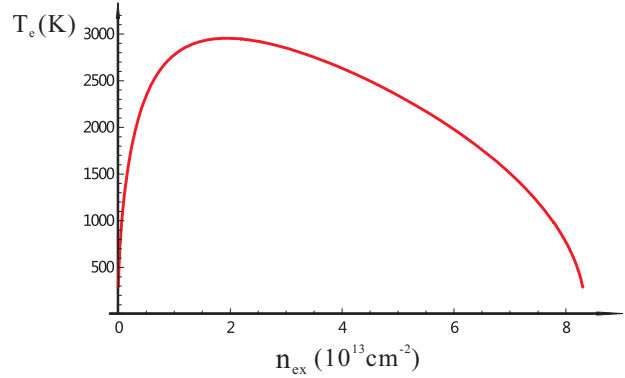}\hfill{}

\caption{Plot of the electron temperature $T_{e}$ varying with photoexcitation
density $n_{\text{ex}}$ in epitaxial graphene on SiC substrate with
initial chemical potential $\mu^{0}=0.4$eV. The non-monotonous behavior
remains although the change of temperature with density is smoother
compared to the neutral system. \label{fig: Expt system T(n_ex)}}
\end{figure}

To show the numerical results, we choose the experimental system of
graphene on SiC substrate with an initial electron doping $\delta=1.17\times10^{13}\text{cm}^{-2}$,
corresponding to an initial chemical potential $\mu^{0}=0.4$eV, being
excited by the pump energy $\hbar\omega=1.55\text{eV}$. The phase
space capacity is calculated from Eq. (\ref{eq: fmax vs x}) to be
$n_{\text{ex}}^{\text{max}}=8.34\times10^{13}\text{cm}^{-2}$ at this
doping level. The electron temperature $T_{e}$ (in units of Kelvin)
changing with $n_{\text{ex}}$ (in units of $10^{13}\text{cm}^{-2}$)
is plotted in Fig.\,\ref{fig: Expt system T(n_ex)}. Compared to
the undoped system, the evolution of electron temperature with photoexcitation
density is smoother and the electron temperature is lower due to the
finite initial carrier density.

\begin{figure}[H]
\hfill{}\includegraphics[scale=0.8]{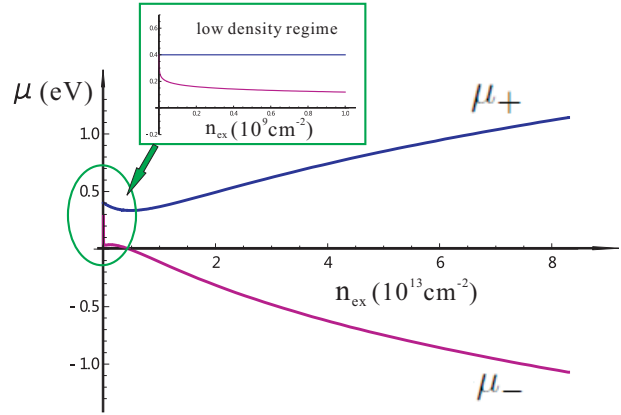}\hfill{}

\caption{Plot of the upper- and lower-band chemical potentials $\mu_{+}$ and
$\mu_{-}$, in units of eV, varying with photoexcitation density $n_{\text{ex}}$
in epitaxial graphene on SiC substrate with initial chemical potential
$\mu^{0}=0.4$eV. The upper-band chemical potential remains positive
although it drops a little bit at low densities, while the lower-band
chemical potential drops rapidly at rather low densities followed
by increasing separation from the upper-band chemical potential. Clearly,
the upper- and lower-band chemical potentials are not symmetric as
in the neutral system. \label{fig: Expt system Mu(n_ex)}}
\end{figure}

Figure \ref{fig: Expt system Mu(n_ex)} shows the upper- and lower-band
chemical potential $\mu_{+}$ and $\mu_{-}$ (in units of eV) varying
with $n_{\text{ex}}$ (in units of $10^{13}\text{cm}^{-2}$).

Clearly, due to the large initial electron-doping, the low-density
classical gas phase for the upper-band electrons is absent now, although
there is a tiny presence for the lower band. And in the doped case,
the upper- and lower-band chemical potentials are not symmetric, $\mu_{+}\neq-\mu_{-}$,
as they are in the neutral system. The separation between the two
chemical potentials increases with photoexcitation density.

\subsubsection{Broadband Distribution and Blueshifted Photoluminescence}

\begin{figure}[H]
\hfill{}\includegraphics[scale=0.8]{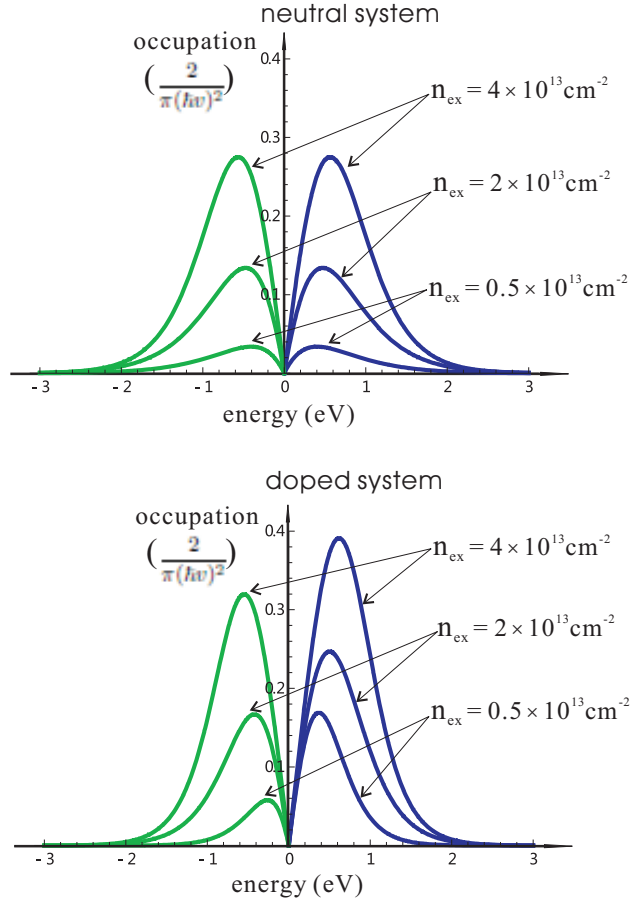}\hfill{}

\caption{Electron (blue line) and hole (green line) occupation number, in units
of $\frac{2}{\pi(\hbar v)^{2}}$, distributed with respect to energy
at different photoexcitation density states. We can see the broadband
distribution of photoexcited carriers with a high-temperature tail
up to almost 2 eV. \label{fig: e and h occupation}}
\end{figure}

A direct consequence to the high electron temperature $T_{e}\sim3000-4000\text{K}$
and the slow population relaxation is a broadband distribution of
electron and hole excitations. This can be illustrated in the occupation
number $N_{e,h}(\varepsilon)=D(\varepsilon)f_{e,h}(\varepsilon)$,
where $D(\varepsilon)=\frac{2\varepsilon}{\pi(\hbar v)^{2}}$ is the
density of state at energy $\varepsilon$ and $f_{e,h}(\varepsilon)$
are the electron and hole distribution functions with $f_{e}=f_{+},\ f_{h}=1-f_{-}$.
Figure \ref{fig: e and h occupation} shows the electron and hole
distribution at different photoexcitation density in the neutral system
and in the electron-doped system ($\mu^{0}=0.4\text{eV}$).

The high temperature tail in the distribution extends the excited
carriers well above the excitation energy $1.55\text{eV}$ up to 2-3eV.
This coverage of higher energy states enables emission of photons
with higher frequencies than that of the excitation photons, which
exhibits blue-shift phenomena in the photoluminescence spectrum. This
unusual blueshifted components have been observed in recent experiments\cite{Lui10,Liu10,Stohr10}
and is naturally explained in our model.

\subsection{Optical Conductivity}

Now we are back to the question raised before. In order to identify
the saturation density, we need to study the optical responses of
this electronic state at different photoexcitation densities. In this
section, we calculate the optical conductivity for the nonequilibrium
intermediate state using Keldysh technique. By analyzing its behavior
at high densities, interesting optical properties are revealed.

\subsubsection{General Formalism}

The low-energy noninteracting Hamiltonian of graphene can be written
in the band representation as 
\begin{equation}
H_{0}=v\sum_{a=1}^{g}\int_{\mathbf{k}}\sum_{\lambda=\pm}\lambda k\gamma_{a,\lambda}^{\dagger}(\mathbf{k})\gamma_{a,\lambda}(\mathbf{k})
\end{equation}
with $\lambda=+$ or $-$ corresponding to the upper or lower band
and $\gamma_{a,\lambda}^{\dagger}(\mathbf{k}),\ \gamma_{a,\lambda}(\mathbf{k})$
are the operators that create or annihilate a quasiparticle of flavor
$a$ (spin and valley) at the 2-dimensional wavevector $\mathbf{k}=(k_{x},k_{y})$
in band $\lambda$. In the band representation the current vertex
becomes 
\begin{equation}
\hat{\mathbf{j}}_{\mathbf{k}}=ev\left(\frac{\mathbf{k}}{k}\sigma_{z}-\frac{\mathbf{k}\times\mathbf{e}_{z}}{k}\sigma_{y}\right),
\end{equation}
where $\sigma_{y,z}$ are Pauli matrices due to the chiral structure
of the Dirac fermions in graphene. Note that we set $\hbar\equiv1$
during the derivation, but will recover it in the final results.

In Keldysh formalism, the bubble diagram contributing to the optical
conductivity gives the real part as 
\begin{equation}
\mathrm{Re}\sigma_{\alpha\beta}(\omega)=\frac{g\pi}{\omega}\int\frac{d\omega'd^{2}k}{(2\pi)^{2}}\mathrm{Tr}\left[\hat{j}_{\alpha\mathbf{k}}\hat{A}_{\mathbf{k}}(\omega'+\omega)\hat{j}_{\beta\mathbf{k}}\hat{N}_{\mathbf{k}}(\omega')-\hat{j}_{\alpha\mathbf{k}}\hat{A}_{\mathbf{k}}(\omega')\hat{j}_{\beta\mathbf{k}}\hat{N}_{\mathbf{k}}(\omega'+\omega)\right]\label{eq: conductivity}
\end{equation}
for the direction $\alpha(\beta)=x,y$ in the 2D graphene layer with
the definitions 
\begin{eqnarray}
\hat{A}_{\mathbf{k}}(\omega) & = & \frac{i}{2\pi}\left(\hat{G}_{\mathbf{k}}^{ret}(\omega)-\hat{G}_{\mathbf{k}}^{adv}(\omega)\right),\label{eq: Sepctrual function}\\
\hat{N}_{\mathbf{k}}(\omega) & = & -\frac{i}{2\pi}\hat{G}_{\mathbf{k}}^{<}(\omega).\label{eq: Lesser function}
\end{eqnarray}
Here the retarded and advanced Green's functions are matrices in band
representation 
\begin{equation}
\hat{G}_{\mathbf{k}}^{ret/adv}(\omega)=\left(\begin{array}{cc}
\frac{1}{\omega\pm i0^{+}+\bar{\mu}-vk} & 0\\
0 & \frac{1}{\omega\pm i0^{+}+\bar{\mu}+vk}
\end{array}\right)\label{eq: retarded/advanced Green}
\end{equation}
where $+$ ($-$) sign associates with the retarded (advanced) Green's
function, and the lesser Green's function is written as 
\begin{equation}
\hat{G}_{\mathbf{k}}^{<}(\omega)=\left(\begin{array}{cc}
g_{\mathbf{k},+}^{<}(\omega) & 0\\
0 & g_{\mathbf{k},-}^{<}(\omega)
\end{array}\right),
\end{equation}

\begin{equation}
g_{\mathbf{k},\pm}^{<}(\omega)=2\pi if(\varepsilon_{\pm}(k)-\mu_{\pm})\delta(\omega-\varepsilon_{\pm}(k)+\bar{\mu}),\label{eq: lesser Green}
\end{equation}
with $\bar{\mu}=\frac{1}{2}\left(\mu_{+}+\mu_{-}\right)$ and the
Fermi function $f(x)\equiv\frac{1}{e^{x/(k_{\text{B}}T_{e})}+1}.$
Note that distinct chemical potentials are employed in the distribution
functions to characterize the nonequilibrium state but an average
chemical potential is used in the spectral functions to avoid an artificial
modification of the spectrum.

Previous analysis (see Section 3) shows that the optical properties
of graphene on an insulating substrate, such as reflection, transmission,
and absorption, are fully determined by the real part of the optical
conductivity $\mathrm{Re}\sigma(\omega)$ to leading order in the
fine-structure constant of quantum electrodynamics $\alpha_{\text{QED}}=\frac{e^{2}}{\hbar c}\approx\frac{1}{137}\ll1$.
The imaginary part only enters at higher orders. Therefore, we make
direct connections of the real part conductivity to the optical responses
observable in experiments.

From Eq.(\ref{eq: conductivity}) we calculate the longitudinal conductivity
which contains intraband and interband transitions. To show the transition
processes specifically, introduce $a_{\mathbf{k},\lambda}(\omega)\equiv\delta(\omega-\lambda vk+\bar{\mu}),\ \text{and}\ f_{\lambda}(\omega)\equiv f(\omega-\lambda\delta\mu).$
It follows the intraband and interband conductivities 
\begin{eqnarray}
 &  & \mathrm{Re}\sigma_{xx}^{\text{intra}}(\omega)\nonumber \\
 & = & g\pi\left(ev\right)^{2}\int\frac{d\omega'd^{2}k}{(2\pi)^{2}}\cos^{2}\theta\sum_{\lambda=\pm}a_{\mathbf{k},\lambda}(\omega')a_{\mathbf{k},\lambda}(\omega'+\omega)\frac{f_{\lambda}(\omega')-f_{\lambda}(\omega'+\omega)}{\omega},\label{eq: intra-band conductivity}\\
 &  & \mathrm{Re}\sigma_{xx}^{\text{inter}}(\omega)\nonumber \\
 & = & g\pi\left(ev\right)^{2}\int\frac{d\omega'd^{2}k}{(2\pi)^{2}}\sin^{2}\theta\sum_{\lambda=\pm}a_{\mathbf{k},\lambda}(\omega')a_{\mathbf{k},\bar{\lambda}}(\omega'+\omega)\frac{f_{\lambda}(\omega')-f_{\bar{\lambda}}(\omega'+\omega)}{\omega},\label{eq: inter-band conductivity}
\end{eqnarray}
where $\cos\theta=\frac{k_{x}}{k},\ \sin\theta=\frac{k_{y}}{k}$.

\subsubsection{Intraband Transition}

First let us evaluate intraband conductivity. It is straightforwardly
obtained from Eq. (\ref{eq: intra-band conductivity}) 
\begin{eqnarray}
\mathrm{Re}\sigma_{xx}^{\text{intra}}(\omega) & = & \frac{g\left(ev\right)^{2}}{(2)^{2}}\delta(\omega)\int_{0}^{\infty}kdk\left[-\frac{\partial f(\omega)}{\partial\omega}\Biggl|_{\omega=vk-\mu_{+}}-\frac{\partial f(\omega)}{\partial\omega}\Biggl|_{\omega=-vk-\mu_{-}}\right]\nonumber \\
 & = & \frac{e^{2}}{\hbar}\ln\left[(1+z_{+})(1+z_{-}^{-1})\right]k_{B}T_{e}\delta(\omega)\label{eq: intra-band formula}
\end{eqnarray}
where we have recovered the factor $\hbar$ on the last line. The
delta-function will be replaced by a Lorentzian $\delta(\omega)\rightarrow\frac{\tau^{-1}}{\omega^{2}+(\tau^{-1})^{2}}$
for further discussions, which is not our concern here. In equilibrium
state $z_{+}=z_{-}=z^{0}=e^{\beta\mu^{0}}$, the intraband conductivity
becomes $\mathrm{Re}\sigma_{xx}^{\text{intra}}(\omega)=\frac{e^{2}}{\hbar}\left[k_{\text{B}}T_{e}^{0}\ln(2+e^{\beta\mu^{0}}+e^{-\beta\mu^{0}})\right]\delta(\omega)$
which recovers the well-known expression in the neutral system at
equilibrium $\mathrm{Re}\sigma_{xx}^{\text{intra}}(\omega)=2\ln2\frac{e^{2}}{\hbar}k_{\text{B}}T_{e}^{0}\delta(\omega),$
as expected.

\begin{figure}[H]
\hfill{}\includegraphics[scale=0.65]{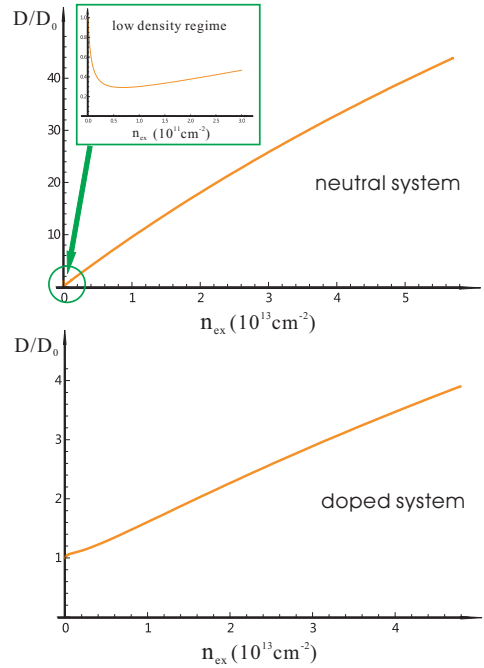}\hfill{}

\caption{Drude spectral weigh $D$, normalized by the equilibrium value $D_{0}$,
changes with photoexcitation density in neutral and doped (with initial
chemical potential $\mu^{0}=0.4\text{eV}$) system. In neutral system
it drops at low densities due to the large drop in chemical potential
as shown in Fig.\,\ref{fig: neutral system MU(n_ex)}, but followed
by a large enhancement at higher densities. In the doped system, it
is always enhanced but with a much smaller enhancement factor than
that of the neutral system. \label{fig:Drude-spectrual-weigh}}
\end{figure}

The most interesting observation from the intraband transition for
the transient electronic state is the modified Drude spectral weight
\begin{equation}
D=\frac{e^{2}}{\hbar}\ln\left[(1+z_{+})(1+z_{-}^{-1})\right]k_{B}T_{e},\label{eq: Drude weight}
\end{equation}
which can be significantly enhanced by the high electron temperature.
But it could also be reduced when the reduction of chemical potential
dominates at low densities in the neutral system. Here we show the
Drude weight change, normalized by the equilibrium value $D_{0}=\frac{e^{2}}{\hbar}\ln\left[(1+z^{0})(1+1/z^{0})\right]k_{B}T_{e}^{0}$,
at different $n_{\text{ex}}$ in the neutral and electron-doped systems
in Fig. \ref{fig:Drude-spectrual-weigh}.

The results for the neutral system exhibits a drop in Drude spectral
weight at low densities due to the large drop in chemical potential
as shown in Fig.\,\ref{fig: neutral system MU(n_ex)} but quickly
followed by large enhancement at higher densities. In the electron-doped
system, Drude weight is always increasing but with much less enhancement
than in the neutral system. 

\subsubsection{Interband Transition}

In order to understand the optical response at high frequencies, as
optical conductivity is dominated by interband transition for frequencies
on the order of 1eV, we evaluate the interband conductivity from Eq.(\ref{eq: inter-band conductivity}),
\begin{eqnarray}
\mathrm{Re}\sigma_{xx}^{\text{inter}}(\omega) & = & \frac{g\left(e\right)^{2}}{(2)^{4}}\left[f(-\frac{\omega}{2}-\mu_{-})-f(\frac{\omega}{2}-\mu_{+})\right]\nonumber \\
 & = & \frac{e^{2}}{4\hbar}\frac{1}{2}\left[\tanh\left(\frac{\hbar\omega+2\mu_{-}}{4k_{B}T_{e}}\right)+\tanh\left(\frac{\hbar\omega-2\mu_{+}}{4k_{B}T_{e}}\right)\right]\label{eq: inter-band formula}
\end{eqnarray}
where the probe photon frequency $\omega>0$ and we have reinserted
the factor $\hbar$ on the last line. In equilibrium state, $\mu_{+}=\mu_{-}=\mu^{0}$,
the interband transition becomes $\mathrm{Re}\sigma_{xx}^{\text{inter}}(\omega)=\frac{e^{2}}{4\hbar}\frac{1}{2}\left[\tanh\left(\frac{\hbar\omega+2\mu^{0}}{4k_{B}T_{e}^{0}}\right)+\tanh\left(\frac{\hbar\omega-2\mu^{0}}{4k_{B}T_{e}^{0}}\right)\right]$
which gives the expression for the neutral system in equilibrium $\mathrm{Re}\sigma_{xx}^{\text{inter}}(\omega)=\frac{e^{2}}{4\hbar}\tanh\left(\frac{\hbar\omega}{4k_{B}T_{e}^{0}}\right)$,
as expected.

By studying the optical response to different photon energies at various
photoexcitation densities, unusual optical properties of the transient
electronic state are found, which will be discussed in the following.

\paragraph{Femtosecond Absorption Saturation and Perfect Transparency}

Let us first consider the optical response to the pump frequency.
An interesting observation from the interband transition formula (\ref{eq: inter-band formula})
arises due to the two density-dependent chemical potentials. As shown
in Fig.\ref{fig: neutral system MU(n_ex)} and \ref{fig: Expt system Mu(n_ex)},
with increasing photoexcitation density, the separation of the two
chemical potentials also gets larger. Then at a certain photoexcitation
density such that 
\begin{equation}
\mu_{+}-\mu_{-}=\hbar\omega,
\end{equation}
the optical conductivity vanishes and the system becomes perfect transparent.
For higher densities, $\mu_{+}-\mu_{-}>\hbar\omega$ such that optical
conductivity turns into negative, which implies a stimulated emission
to keep the photoexcitation density from rising. This indicates that
the absorption reaches zero and the number of excited carriers saturates
at this density $n_{\text{ex}}^{\text{sat}}$ , which is stabilized
by stimulated emission.

\begin{figure}[H]
\hfill{}\includegraphics[scale=0.73]{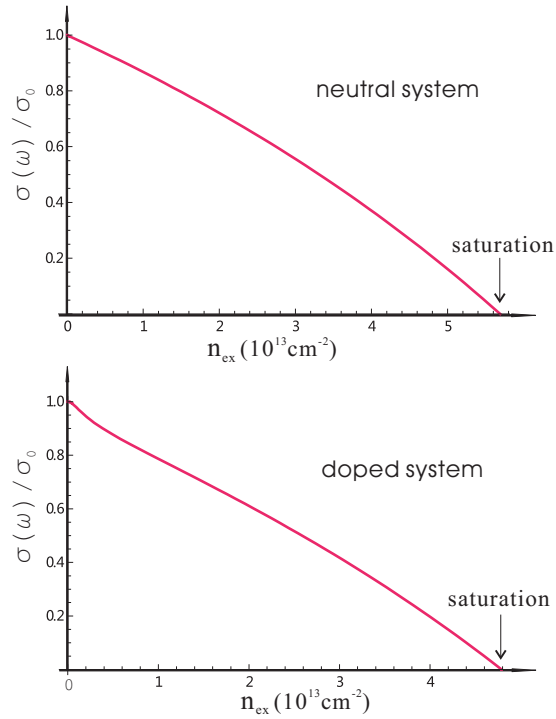}\hfill{}

\caption{Interband conductivity $\sigma(\hbar\omega=1.55\text{eV})$, normalized
by the equilibrium value $\sigma_{0}=\frac{e^{2}}{4\hbar}$, varies
with photoexcitation density $n_{\text{ex}}$ in the neutral and doped
systems. It reaches zero, which indicates complete bleaching of absorption,
at $5.7\times10^{13}\text{cm}^{-2}$ in the neutral system and $4.8\times10^{13}\text{cm}^{-2}$
in the doped system, corresponding to the saturation densities. \label{fig: Sigma of n_ex}}
\end{figure}

To show the variation of optical conductivity with photoexcitation
density, we again calculate for the neutral system and the electron-doped
system with initial doping $\mu^{0}=0.4$eV by applying pump photon
energy at $\hbar\omega=1.55\text{eV}$. We show the results in Fig.\ref{fig: Sigma of n_ex}
where they are normalized by the equilibrium value $\sigma_{0}=\frac{e^{2}}{4\hbar}$.

In both cases, the conductivity monotonously decreases due to the
increasing electron temperature and separation of chemical potentials.
The neutral system saturates at $n_{\text{ex}}^{\text{sat}}(\text{theory})=5.7\times10^{13}\text{cm}^{-2}$
while the electron-doped system saturates at roughly $n_{\text{ex}}^{\text{sat}}(\text{theory})=4.8\times10^{13}\text{cm}^{-2}$.
On the other hand, experimental measurement of the electron-doped
system gives $n_{\text{ex}}^{\text{sat}}(\text{expt.})=5.0\times10^{13}\text{cm}^{-2}$,
as shown in Fig. 6b, which is in excellent agreement with the theoretical
value $4.8\times10^{13}\text{cm}^{-2}$. This corroborates the correct
description of the transient electronic state in our model. And it
also answers the early posted question: the system saturates at a
lower photoexcitation density before completely filling the available
phase space, i.e., $n_{\text{ex}}^{\text{sat}}<n_{\text{ex}}^{\text{max}}$.

\begin{figure}[H]
\hfill{}\includegraphics[scale=0.73]{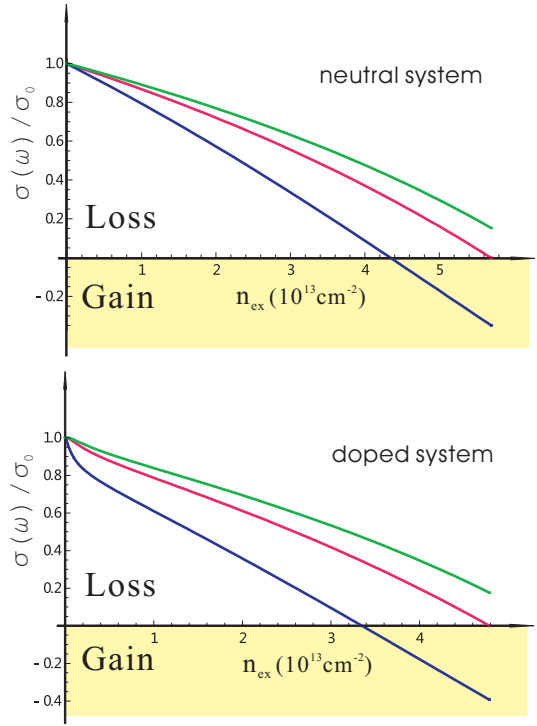}\hfill{}

\caption{The calculated interband conductivity $\sigma(\omega)$, normalized
by the equilibrium value $\sigma_{0}=\frac{e^{2}}{4\hbar}$, at higher
($\hbar\omega=1.7\text{eV}$, green line), the pump ($\hbar\omega=1.55\text{eV}$,
magenta line), and lower ($\hbar\omega=1.2\text{eV}$, blue line)
frequency. It shows that in the vicinity of saturation density the
optical conductivity at lower frequencies (but still high enough to
mainly detect interband transition) becomes negative. This indicates
a stimulated emission that enables an optical gain for the low-frequency
probes. \label{fig:Negative-optical-conductivity}}
\end{figure}

This theoretical calculation of an optical gain can serve as a test
of our model, which have been used to simulate the optical differential
reflectivity data performed in Fig. 5b (red line). The agreement is
excellent.

\paragraph{Femtosecond Stimulated Emission and Optical Gain}

It is easy to see that higher-frequency pump will saturate at higher
density since it can open up more phase space. Then when applying
a probe with frequency higher than the pump frequency, we will expect
it can not detect the zero absorption, as long as it is still within
the low-energy Dirac spectrum, as shown in Fig.\ref{fig:Negative-optical-conductivity}
(green line). However, if one applies a lower-frequency probe, but
not too low such that it is still mainly detecting the change in interband
transition, one would expect an optical gain in the vicinity of the
saturation density, as the optical conductivity becomes negative in
this regime as shown in Fig.\ref{fig:Negative-optical-conductivity}
(blue line, yellow region). The appearance of negative conductivity
signifies a stimulated emission that drives the system to a lower
density as illustrated in Fig.\ref{fig:stimulated emission}. We stress
that the transient conductivity is negative in a regime below the
pump frequency but above a certain frequency below which intraband
transition becomes dominant.

\paragraph{Comparison of Two Model Calculations with Pump-probe Spectroscopy
Measurements}

\begin{figure}[H]
\hfill{}\includegraphics[scale=0.5]{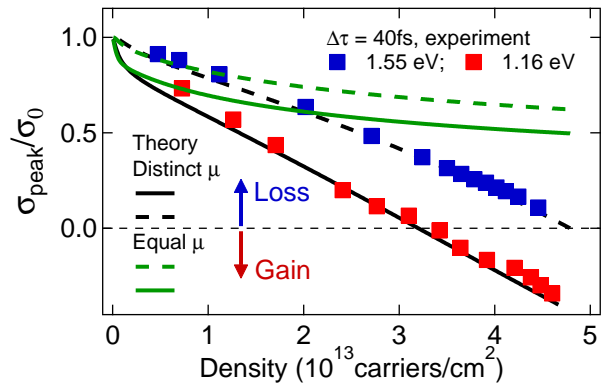}\hfill{}

\caption{Comparison of the calculation from the distinct-$\mu$ model ($\mu_{+}\neq\mu_{-}$)
and the equal-$\mu$ model ($\mu_{+}=\mu_{-}$) with the experimentally
measured transient optical conductivity at 40 fs after the 1.55eV
pump at varying pump fluence. The probe photon energies are $\hbar\omega=1.55\text{eV}$
(blue solid square) and $1.16\text{eV}$ (red solid square). Clearly,
the distinct-$\mu$ model calculation of the conductivity at 1.55eV
(black dashed line) and 1.16eV (black solid line) agrees quantitatively
with the experimental data, in sharp contrast to the equal-$\mu$
model results (green dashed line for 1.55eV and green solid line for
1.16eV). For the probe photon at lower frequency 1.16eV in the non-degenerate
scheme, the transient conductivity becomes negative above a critical
density exhibiting a transition from optical loss to gain, predicted
by distinct-$\mu$ model and substantiated by experiment. \label{fig:models comparison with expt} }
\end{figure}

Here we further compare the calculated optical conductivity from the
distinct-$\mu$ model ($\mu_{+}\neq\mu_{-}$) discussed above and
the equal-$\mu$ model ($\mu_{+}=\mu_{-}$) with the experimental
value measured at probe photon energy $\hbar\omega=1.55\text{eV}\ \text{and}\ 1.16\text{eV}$
for a pump energy at $\hbar\omega=1.55\text{eV}$. As shown in Fig.\ref{fig:models comparison with expt},
we compare the experimentally-extracted, transient conductivity at
40 fs \cite{TianqiLi10} with the calculated conductivity $\sigma(\omega)$
as a function of the photoexcited carrier density $n_{\text{ex}}$
for two probe photon energies 1.55 eV and 1.16 eV. The Fermi energy
of the sample is $\sim0.4$ eV. The model calculation with the distinct
chemical potentials reproduces the salient features of the experiment
including nonlinear saturation and optical gain. Excellent agreement
between experiment and theory also demonstrates a faithful representation
of the transient state at 40 fs by the model described in the manuscript.
The model calculation with the same chemical potential clearly fails
to account for the experimental observations. For the degenerate scheme,
our theory (black dashed line) yields $\sigma\rightarrow0$ and thus
perfect transparency at $n_{\text{ex}}=4.8\times10^{13}\text{cm}^{-2}$.
Once the system is driven into this regime, a balance between stimulated
emission and absorption will lead to a transparency. For non-degenerate
scheme by probing at 1.16 eV, our theory (black solid line) predicts
a critical density $3.2\times10^{13}\text{cm}^{-2}$ for the transition
from loss to gain. All of these results agree quantitatively with
the experimental values $5.0\times10^{13}\text{cm}^{-2}$ and $3.4\times10^{13}\text{cm}^{-2}$,
respectively.

\section{Conclusions}

We have studied the electronic state in photoexcited graphene formed
via rapid carrier-carrier scattering after strong photoexcitation
but before energy relaxation that takes place on a longer time scale.
We have provided evidence for the existence of pronounced femtosecond
population inversion and broadband gain in strongly photoexcited graphene
monolayers. These results clearly reveal the transient electron and
hole potentials are separated on the time scale of 100s of fs. By
characterizing the state in terms of two separate Fermi-Dirac distributions
with a common electron temperature but distinct chemical potentials
for the upper and lower bands, we showed that this intermediate electronic
state is associated with high electron temperature $T_{e}$ up to
3000-4000K, which causes a broadband distribution extended to higher
energy and is responsible for the observed blueshifted photoluminescence
component. The analysis on the variation of electron temperature and
chemical potentials with photoexcited carrier density in the neutral
system clearly shows a crossover from hot dilute classical gas to
dense quantum degenerate fermions. And unlike the phase space restriction
in most semiconductors for a pump pulse on the order of 10fs, which
is determined by the density of state at the optical excitation and
the frequency width of the pulse, the fast depletion of phase space
in graphene yields a broadband filling which significantly enlarge
the accommodation of photoexcited carriers.

\ack{}{}

We thank Myron Hupalo and Michael Tringides for discussions. J.Z.
acknowledges support by the Jeffress Memorial Trust, Grant No. J-1033.
J.S. thanks the DFG Center for Functional Nanostructures. J.W. and
T.L. acknowledge support by the the National Science Foundation (contract
no. DMR-1055352). Work at Ames Laboratory was partially supported
by the U.S. Department of Energy, Office of Basic Energy Science,
Division of Materials Sciences and Engineering (Ames Laboratory is
operated for the U.S. Department of Energy by Iowa State University
under Contract No. DE-AC02-07CH11358).

\bigskip{}

\bigskip{}

\bigskip{}

\end{document}